\documentstyle[11pt]{article}
\input{epsf}
\def\la{\;
\raise0.3ex\hbox{$<$\kern-0.75em\raise-1.1ex\hbox{$\sim$}}\; }
\def\ga{\;
\raise0.3ex\hbox{$>$\kern-0.75em\raise-1.1ex\hbox{$\sim$}}\; }
\newcommand{\om}{\mbox{$\omega$}}              % \omega
\newcommand{\dd}{\mbox{d}}                     % d - differential
\newcommand{\vp}{\mbox{\boldmath $p$}}         % vector p (momentum)
\newcommand{\vk}{\mbox{\boldmath $k$}}         % vector k (phonon)
\newcommand{\vq}{\mbox{\boldmath $q$}}         % vector q (momentum)
\newcommand{\kB}{\mbox{$k_{\rm B}$}}           % Boltzmann constant
\newcommand{\vect}[1]{\mbox{\boldmath $#1$}}   % vector function
\begin{document}

\title{Neutrino-pair emission due to electron-phonon scattering
      in a neutron star crust}

\author{
        D.G. Yakovlev and A.D. Kaminker \\
        Ioffe Physical-Technical Institute \\
        194021, St.Petersburg, Russia}
\date{}
\maketitle

% Abstract ********************************************************
\begin{abstract}

Neutrino-pair bremsstrahlung radiation is considered due
to electron--phonon scattering
of degenerate, relativistic electrons in a
lattice of spherical atomic nuclei in a neutron star
crust. The neutrino energy generation rate is calculated
taking into account exact spectrum of phonons, the
Debye--Waller factor, and the nuclear form--factor in the
density range from
$10^7$~g~cm$^{-3}$
to $10^{14}$~g~cm$^{-3}$
at arbitrary nuclear composition
for body-centered-cubic and face-centered-cubic Coulomb crystals.
The results are fitted by a unified analytic expression.
A comparison is given of the neutrino bremsstrahlung energy losses
in a neutron star crust composed of ground state and accreted matter,
in the solid and liquid phases.
\end{abstract}

% Sec.1 ************************************************************
\section{Introduction}
% ******************************************************************
The neutrino-pair bremsstrahlung radiation of electrons in
the liquid and crystalline phases of dense matter is one
of the main cooling mechanisms of neutron star (NS) crusts.
This radiation has been studied in a number of works
(see, e.g., Itoh et al., 1989, and references therein).
Recently Haensel et al. (1966) have reconsidered
the neutrino radiation produced due to collisions
of degenerate, relativistic electrons with atomic nuclei
in the liquid phase,
$ e + (Z,A) \to e + (Z,A) + \nu + \bar{\nu}$.
The authors have expressed the neutrino energy loss rate
in a simple form through the Coulomb
logarithm which depends weakly on the parameters
of stellar matter. The Coulomb logarithm has been calculated and
fitted analytically for matter with arbitrary nuclear composition
at all densities and temperatures of practical interest.
The proposed approach is invalid only in the deep layers of
the NS crust ($\rho \sim 10^{14}$ g~cm$^{-3}$),
where the atomic nuclei can form clusters and/or have nonspherical shape
(Lorenz et al., 1993, Pethick and Ravenhall, 1995).

In the present article, we also restrict ourselves to consideration
of spherical atomic nuclei. We will study the neutrino-pair
bremsstrahlung of electrons in the crystalline phase of dense
matter. It is well known (e.g., Itoh et al., 1989)
that this process has two channels.
The first one is the neutrino radiation due to Bragg diffraction
of electrons in a lattice
(``the static lattice contribution'').
The second one is the neutrino emission due to absorption or creation
of phonons (``the phonon contribution'').
It has been accepted until recently that both components
have comparable intensities. However recently Pethick and Thorsson
(1994) have shown that the band structure of crystals in a NS crust
(the presence of an energy gap in the electron dispersion
relation at the Brillouin zone boundaries)
suppresses exponentially the static lattice contribution.
Accordingly the phonon contribution dominates, and it will be the subject
of our consideration.

The neutrino radiation due to the electron--phonon scattering
has been studied earlier by Flowers (1973) and --- in
the frame of the Weinberg -- Salam -- Glashow theory ---
by Itoh et al. (1984a). The latter authors fitted their
results by complicated analytic expressions and tabulated
the fit parameters for matter with different nuclear compositions.
These results are also included in the new review of
Itoh et al. (1996). We will show that, in some cases,
the fits of Itoh et al. (1984a, 1996) are not too accurate.
We will carry out more accurate calculations and fit the
results by a unified formula valid for matter with arbitrary
nuclear composition.
In addition to the body-centered-cubic crystals,
which have been studied
traditionally in the literature, we will consider also the
face-centered-cubic crystals. For illustration, we analyse
the neutrino--pair bremsstrahlung
in the solid and liquid phases of NS crusts composed
of ground--state or accreted matters.

% Sec. 2 *******************************************************
\section{Physical conditions}
% **************************************************************
Consider matter of a NS crust in the density range
$10^6$ g~cm$^{-3}3 \ll \rho \la 10^{14}$ g~cm$^{-3}$.
In the domain of $10^6$ g~cm$^{-3} \ll \rho < \rho_{\rm ND}
\simeq 4-6 \times 10^{11}$ g~cm$^{-3}$,
which corresponds to the outer crust,
the matter consists of atomic nuclei immersed in a nearly ideal
relativistic and degenerate gas of electrons.
In the inner crust, where $\rho > \rho_{\rm ND}$,
free neutrons appear in addition to the nuclei and electrons.
If $\rho \ga 10^{14} $g~cm$^{-3}$
the nuclei can be nonspherical and form clusters
(Lorenz et al., 1993; Pethick and Ravenhall, 1995);
the nuclei disappear
at the bottom of the inner crust, at
$ \rho > 1.5 \times 10^{14}$ g~cm$^{-3}$.

The state of degenerate electrons in the NS crust
is characterized by the Fermi momentum
$p_{\rm F}$ or the relativistic parameter $x$:
\begin{equation}
       p_{\rm F} = \hbar (3 \pi^2 n_{\rm e})^{1/3}, \hspace{5mm}
       x = \frac{p_{\rm F}}{m_e c} \approx 100.9
       \left( \frac{\rho_{12}}{\mu_e} \right)^{1/3},
\label{1}
\end{equation}
where $\mu_e$ is the number of baryons per electron and
$\rho_{12}$ is density in units of $10^{12}$ g~cm$^{-3}$.
In the density range of study we have
$\rho \gg 10^6$ g~cm$^{-3}$, and the electron gas is ultrarelativistic
($x \gg 1$). The electron degeneracy temperature is
\begin{equation}
       T_{\rm F} = T_0 \, (\sqrt{1+x^2}-1), \; \;
             T_0 = \frac{m_e c^2}{k_{\rm B}}
             \approx 5.930 \times 10^9 \;{\rm K},
\label{2}
\end{equation}
where $k_{\rm B}$ is the Boltzmann constant.
We consider the case of $T \ll T_{\rm F}$.

The state of nuclei (ions) is characterized by the
ion--coupling parameter
$ \Gamma = Z^2 e^2/(a k_{\rm B} T)
\approx 0.225 \, x \, Z^{5/3}/T_8$,
where $Ze$ is the nuclear charge, $a=[3/(4 \pi n_{\rm i})]^{1/3}$
is the mean inter--ion distance,
$n_{\rm i}$ is the number density of nuclei, and
$T_8$ is temperature in units of $10^8$ K.
At high enough temperatures, when $\Gamma \ll 1$,
the nuclei constitute the Boltzmann gas.
For lower $T$ ($\Gamma \ga 1$), the gas transforms into the
Coulomb liquid, and the liquid solidifies into the Coulomb
crystal at $T \leq T_{\rm m}$.
The melting temperature of a classical one--component plasma
of ions corresponds to
$\Gamma_{\rm m} = 172$ (Nagara et al., 1987) and equals
\begin{equation}
         T_{\rm m} = \frac{Z^2 e^2}{a k_{\rm B} \Gamma_{\rm m}}
         \,\approx 1.32 \times 10^7 Z^{5/3} \left(
         \frac{\rho_{12}}{\mu_e} \right)^{1/3} {\rm K}.
\label{3}
\end{equation}
Note that the crystallization of light nuclei
(H, He) is additionally suppressed
by strong zero-point vibrations
(Mochkovitch and Hansen, 1979).  The most bound is
the body-centered-cubic (bcc) crystal
(see, e.g., Brush et al., 1966). The binding
of the face-centered-cubic (fcc) crystal is weaker
but very slightly. One cannot exclude that the fcc crystals
appear in dense stellar matter along with the bcc ones
(e.g., Baiko and Yakovlev, 1995). We will consider the temperature
range of crystalline crust $T_{\rm U} < T < T_{\rm m}$,
where $ T_{\rm U} \sim T_{\rm p} Z^{1/3} e^2/(\hbar v_{\rm F})$
is the temperature below which the Umklapp processes are frozen out
(e.g., Raikh and Yakovlev, 1982),
$T_{\rm p}=\hbar \om_{\rm p}/k_{\rm B} \approx 7.83
\times 10^9 \sqrt{ Z \rho_{12}/(A \mu_e)}$~K
is the ion plasma temperature,
$\om_{\rm p} = \sqrt{4 \pi Z^2 e^2 n_{\rm i}/m_{\rm i}}$ is
the ion plasma frequency, and $m_{\rm i}$ is the ion mass.
Note that the Debye temperature of a crystal is
$T_{\rm D} \approx 0.45 T_{\rm p}$ (Carr, 1961).
If $T \ga T_{\rm U}$ the electron--phonon scattering
can be described using the approximation of almost free electrons.
Then, at $Z \gg 1$, the main contribution into the scattering
comes from the Umklapp processes
(Yakovlev and Urpin, 1980, Raikh and Yakovlev, 1982).

For illustration, we will consider two models
of matter in a NS crust: ground state (cold--catalyzed) matter and
accreted matter. The effects of temperature onto the nuclear
composition will be neglected which is justified as long as
$T < 5 \times 10^9$~K (see Haensel et al., 1996).
Both models assume that nuclei of one species are available
in matter at any density (pressure). This leads to jumps of nuclear
composition with increasing density (pressure).

The ground--state matter in the
outer NS crust will be described using the data
obtained by Haensel and Pichon (1994)
from new laboratory measurements of nuclear masses with
large neutron excess. For the inner crust, we will use
the results of Negele and Vautherin (1973) derived on the basis of
a modified Hartree--Fock method.

We will use the recent calculations of Haensel and
Zhdunik (1990b), for describing accreted matter.
The light elements are assumed to be burnt out in the outer
layers of the NS crust while the nuclear composition
in the deeper layers (we are interested in)
is determined by a sequence of nuclear transmutations
which occur in a sinking and compressing accretion matter
and which depend neither on temperature nor on accretion rate.

The comparison of the ground--state and accreted matters
is given, for instance, by Haensel and Zhdunik
(1990a, b) or Haensel et al. (1996).
At $\rho \ga 10^9$ g~cm$^{-3}$, the accreted matter
is generally composed of lighter nuclei with lower $Z$.
For instance, at $\rho = 4 \times 10^{12}$ g~cm$^{-3}$
the accreted matter contains neutron--reach 
magnesium nuclei, $A=44$, $Z=12$
(Haensel and Zhdunik, 1990a, b), whereas the ground--state matter
is composed of tin nuclei, $A=159$, $Z=50$
(Negele and Vautherin, 1973). The above difference is important
for thermodynamic and kinetic properties of matter.
For instance, the melting temperature
(\ref{3}) of the accreted matter at
$10^{11}$ g~cm$^{-3} \la \rho \la 10^{13}$ g~cm$^{-3}$ is about
3 -- 10 times lower than that of the ground--state matter (Figure 1).

% Sec. 3 ***********************************************************
\section{General expressions}
% ******************************************************************
We can use the scheme of extended Brillouin zones and
the free--electron approximation for describing the electron
states. In this case, the Fermi surface and the electron dispersion
relation are the same as for free electrons.
This approximation is violated only in the narrow
regions of momentum space near intersections of the
electron Fermi surface with the Brillouin zone boundaries
where the dispersion relation contains the energy gap.
Note that the interzone electron transitions which
accompany the neutrino--pair emission due to the Bragg
diffraction (Section 1) occur just in these regions.
The presence of the gap suppresses these transitions exponentially
(Pethick and Thorsson, 1994).

A general expression for the neutrino energy loss rate
$Q$ due to the electron--phonon scattering
of relativistic and degenerate electrons
can be obtained using the standard formalism
of the electron--phonon interaction
(Ziman, 1960). In analogy with the results of Haensel et al. (1996),
$Q$ can be written as
\begin{equation}
      Q  =  \frac{8 \pi G_{\rm F}^2 Z^2 e^4 C_+^2}
          {567 \hbar^9 c^8} (k_{\rm B} T)^6 n_{\rm i} L
                 \approx   3.229
         \times 10^{11}\, \rho_{12} \,X_A\,
         \frac{Z^2}{A} \,
         T_8^6 L \; {\rm ergs \; s^{-1} \; cm^{-3}},
\label{4}
\end{equation}
where $G_{\rm F}=1.436 \times 10^{-49}$~ergs~cm$^3$ is the Fermi
weak--coupling constant, $X_A$ is the mass fraction contained in nuclei,
$L$ is a dimensionless function of the parameters of stellar
matter which has meaning of the Coulomb logarithm in the
liquid phase. Furthermore,
$ C_+^2=C_V^2 + C_A^2 + 2({C'}_V^2+ {C'}_A^2)$, 
where $C_V$ and $C_A$ are the vector and axial--vector
constants of weak interaction, respectively. We have
$C_V= 2 \sin^2 \theta_W + 0.5$  and  $C_A= 0.5$,
for the emission of the electron neutrinos
(charged + neutral currents); and
${C'}_V = 2 \sin^2 \theta_W - 0.5$  and  ${C'}_A = -0.5$,
for the emission of the muonic and tauonic neutrinos
(neutral current only).
In this case,  $\theta_W$ is the Weinberg angle,
$\sin^2 \theta_W \simeq 0.23$. Numerical estimate of
$Q$ is obtained taking into account the contribution
of the electron, muon and tau neutrinos which yields
$C_+^2 \approx 1.675$.

The general expression for $L$ in Equation
(\ref{4}) can be presented in the form
\begin{eqnarray}
      L & = & {\alpha_0 \over 2 t S^2} \,
             \int_S \int_S \,
             {\dd S \, \dd S' \over q^2 |\epsilon(q)|^2}
             \left( 1 +  { 2 y^2 \over 1-y^2 } \ln y \right)
             {\rm e}^{-2W(q)}
      \nonumber \\
           & \times &
            |f(q)|^2
            \sum_s [ \vq \vect{e}_s(\vk)]^2
            { H(z) \over z^2},
      \label{5}
\end{eqnarray}
where
\begin{eqnarray}
      t & = & { T \over T_{\rm p}} = 0.0128 T_8
            \left( {A \mu_e \over Z \rho_{12}} \right)^{1/2}, \; \; \;
            z \equiv z_s = {\hbar \omega_s(\vk) \over \kB T},
     \nonumber \\
     H(z) & = & {63 z \over 16 \pi^6}
         \int_0^\infty \, {\rm d}w \, w^4
         \left[ {w+z \over (1 - {\rm e}^{-z})({\rm e}^{w+z}-1)} +
          {w-z \over ({\rm e}^z-1)({\rm e}^{w-z}-1)} \right].
    \label{6}
\end{eqnarray}
In this case, $s$= 1, 2, 3 enumerates three phonon modes
in a Coulomb crystal. The integration in Equation
(\ref{5}) is carried out over possible momenta of electrons
$\vp$  and $\vp'$ on the Fermi surface before and after scattering;
$S = 4 \pi p_{\rm F}^2$ is the Fermi surface area. The quantity
$\hbar \vq = \vp - \vp'$ is an electron momentum transfer in a scattering
event and $y = \hbar q /(2 p_{\rm F})$. Furthermore,
$\vk$, $\omega_s(\vk)$ and $\vect{e}_s(\vk)$ are, respectively,
the wave vector, frequency and polarization unit vector
of a phonon excited or absorbed by an electron:
$\pm \vk = \vq - \vect{K}$, where $\vect{K}$ is an inverse
lattice vector chosen in such a way
that $\vk$ belongs to the first Brillouin zone;
the neutrino--pair momentum can be neglected here.
The case of $\vk = \vq$ ($\vect{K}$=0) corresponds to a so called
normal scattering process, while the case of
$\vk \neq \vq$ ($\vect{K}\neq 0$) corresponds to an Umklapp process.
The quantity $\epsilon(q)$ in Equation (\ref{5}) is
the longitudinal static dielectric function of the electron gas:
it takes into account the electron screening of the Coulomb
potential of a nucleus;
$f(q)$ is the nuclear formfactor to allow for a finite nuclear size.
Two terms under the integral in
$H(z)$ describe the neutrino--pair radiation due to emission and absorption
of phonons. The dimensionless integration variable is
$w= \varepsilon_\nu /(\kB T)$,
where $\varepsilon_\nu$ is the total energy of a neutrino--pair.
Finally, $W(q)$ is the Debye--Waller factor. Baiko and
Yakovlev (1995) fitted this factor for a Coulomb crystal
(with a mean error of about 1\%)
by the expression $2 W(q) = \alpha y^2$, where
\begin{equation}
     \alpha = \alpha_0  \;
             \left( {1 \over 2} u_{-1} {\rm e}^{-9.100t}+tu_{-2} \right),
     \; \; \;
     \alpha_0 = {4 m_e^2c^2 \over k_{\rm B} T_{\rm p} m_{\rm i}} x^2
        \approx 1.683 \sqrt{ { x \over AZ} }.
\label{7}
\end{equation}
Here, $u_n = \langle (\omega / \omega_p)^n \rangle$ is a
frequency moment of the phonon spectrum. Brackets mean
averaging over phonon frequencies and polarizations:
\begin{equation}
     \langle f_s(\vk) \rangle =
     {1 \over 3 V_{\rm B} } \sum_s \int_{V_{\rm B}} \, \dd \vk \,
     f_s(\vk),
\label{8}
\end{equation}
$V_{\rm B}$ is the Brillouin zone volume.
One has $u_{-1}^{(bcc)} = 2.800$,
$u_{-2}^{(bcc)} = 12.998$ for the bcc lattice
(e.g., Mochkovitch and Hansen, 1979), and
$u_{-1}^{(fcc)} = 4.03$,
$u_{-2}^{(fcc)} = 28.8$ for the fcc one (e.g., Baiko and Yakovlev, 1995).
According to Equation (\ref{5}),
the Debye--Waller factor suppresses the electron--phonon
interaction at large momentum transfers $q$,
and weakens the neutrino emission. The factor is important
at large ion vibrations in the lattice --- near the melting
point (where the thermal vibrations are especially strong)
and at high densities (where zero--point vibrations are huge).
Under the conditions of study (Section 2) this factor is
most significant.

In calculations based on Equation (\ref{5})
we will use the static dielectric function of the
degenerate relativistic electron gas
$\epsilon(q)$ derived by Jancovici (1962).
In our case, the electron screening is weak and 
has practically no effect on the results. Nevertheless we
will take it into account.

We will use the nuclear formfactor appropriate
for spherical nuclei with a uniform proton core:
\begin{equation}
       f(q) = \frac{3}{(q r_c)^3}
               \left[ \sin(q r_c) - q r_c \cos(q r_c) \right],
\label{9}
\end{equation}
where, $r_c$ is the core radius (which can be noticeably smaller
than the total nucleus radius if the nucleus is neutron--rich).
The formfactor suppresses the electron--phonon
scattering at large momentum transfers $q$.
One has $f(q)=1$, for point--like nuclei.

It is convenient to fit $H(z)$ by an analytic function.
One can easily see that $H(0)=1$, and $H(z)=21 \, z^7 \,{\rm e}^{-z}/
(160 \pi^6)$ at $z \to \infty$. We have calculated $H(z)$
numerically for intermediate values of $z$. The calculated values and
the asymptotes are reproduced (with the mean error of
1.7 \%) by the fit
\begin{equation}
    H(z) = \left( 0.2968 + \sqrt{0.01633 \, z^2 + 0.7032^2} \right)^7 \;
           \exp \left(5.5032- \sqrt{z^2 + 5.5032^2} \right).
\label{10}
\end{equation}
%

% Sec. 4 *****************************************************
\section{Calculations and fitting of the results}
%*************************************************************
Evaluation of $L$ reduces to a 4-fold integration in Equation
(\ref{5}) over positions of electron momenta on
the Fermi surface before and after scattering. For some
parameters of stellar matter, the integration has been carried
out numerically by the Monte Carlo method. We have used the code
developed by Baiko and Yakovlev (1995) for calculating analogous
integrals which determine the thermal and electric conductivities
of electrons due to electron--phonon scattering. We have modified
the code for calculating $L$. Typical calculation error
for $10^6$ Monte Carlo configurations (choices of positions of electron
momenta on the Fermi surface before and after scattering)
has been 6 \% (at the $3 \sigma$ level).

In order to simplify calculations and analyse the results,
we have used also the simplified method of integration
of the equations similar to (\ref{5}), which was
used earlier by Yakovlev and Urpin (1980), Raikh and Yakovlev (1982),
Itoh et al. (1984b, 1993), and Baiko and Yakovlev (1995) in calculations
of the thermal and electric conductivities. The method is based on
the fact that the main contribution into the electron--phonon scattering
comes from the Umklapp processes. The Fermi
surface is intersected by many boundaries of the Brillouin zones.
Then, for the Umklapp processes, in Equation
(\ref{5}) one can approximately set
\begin{equation}
         \sum_s [ \vq \vect{e}_s(\vk)]^2
          H(z) z^{-2}
        \approx  q^2 t^2 G(t),
\label{11}
\end{equation}
\noindent
\begin{equation}
   G(t)  =   t^{-2}   \left\langle  H(z) z^{-2} \right\rangle,
\label{12}
\end{equation}
where brackets denote the averaging (\ref{8}).
As a result, from Equation (\ref{5}) we obtain
\begin{equation}
       L = \; \alpha_0 t G(t) P.
\label{13}
\end{equation}
Here,
\begin{equation}
       P =
       \int_{y_0}^1 \; y \, \dd y \;
               {|f(q)|^2   \over |\epsilon(q)|^2}
             \left( 1 +  { 2 y^2 \over 1-y^2 } \ln y \right)
               {\rm e}^{- \alpha y^2},
\label{14}
\end{equation}
$y=\hbar q/(2 p_{\rm F})$, and the lower integration limit
$y_0 = \hbar q_0 /(2 p_{\rm F})= (4Z)^{-2/3}$
is chosen in such a way to include the Umklapp processes
but exclude the normal ones.
The minimum momentum $q_0=(6 \pi^2 n_{\rm i})^{1/3}$ is set
equal to the radius of the sphere whose volume equals
the Brillouin zone volume.

Calculation of $L$ from the approximate expression (\ref{13})
reduces to determination of the two functions, $G$ and $P$,
that is much simpler than the exact Monte Carlo integration.
Thus the calculations have been done mainly by the
simplified method whose accuracy has been controlled by the
Monte Carlo runs at some parameter values. In all the cases,
the results have coincided, within the Monte Carlo error bars.

The function $G$ depends on the dimensionless temperature
$t$, given by Equation (\ref{6}), and on the lattice type.
If $t \gg 1$, from Equations (\ref{6}) and (\ref{12})
we obtain the asymptote $G(t) = u_{-2}$, where $u_{-2}$ is a
frequency moment of the phonon spectrum (Section 3).
When $t \to 0$  we have  $G(t) \sim t$. We have calculated
$G(t)$ from Equation (\ref{12}) for the bcc and fcc lattices
using the approximate but highly accurate method proposed by
Mochkovitch and Hansen (1979) for averaging
the expressions similar to (\ref{8})
over the Brillouin zone. The results can be fitted
(with about 1.5 \% mean error) by the function
\begin{equation}
         G(t)= {u_{-2} t \over \sqrt{ a_0^2 + t^2}}
             + {b_1 t^3  \over  (b_2 +t^2)^2},
\label{15}
\end{equation}
where $a_0^{(bcc)} = 0.06423$,  $b_1^{(bcc)} = 0.1151$,
$b_2^{(bcc)} = 5.92 \times 10^{-4}$;
$a_0^{(fcc)} = 0.001479$, $b_1^{(fcc)} = 0.2381$,
$b_2^{(fcc)} = 5.161 \times 10^{-4}$. This expression
reproduces also the above asymptotes.

The function $P$ is given by a simple one--dimensional integral,
which is insensitive to the crystal type
but depends on three parameters:
atomic number $Z$ (through the lower integration limit, $y_0$),
the Debye--Waller parameter $\alpha=\alpha(t)$
(see Equation (\ref{7})), and the parameter
$\eta = r_c/a$ which accounts for the finite size
of atomic nuclei in the nuclear formfactor (\ref{9})
($a$ is given by Equation (\ref{3})).
We have evaluated $P$ from Equation (\ref{14}) at $Z = 20, 40, 60$,
$\alpha$ = 0.04, 0.12, 0.4, 1.2, 4, 12 and
$\eta$ = 0, 0.10, 0.2, 0.3, 0.4.
Analysing the parameters of matter in a NS crust at
$10^6$ g~cm$^{-3} \la \rho \la 10^{14}$ g~cm$^{-3}$
(Section 2), one can show that the selected values
cover all the domain of possible parameter values.
The results are fitted by the expression
\begin{equation}
         P(Z, \alpha, \eta) =
         {F_0 \over 2} \,  \left[ 1+ {(18 \pi Z)^{2/3} \over 5 N}
         {F_1 \over F_0} \eta^2  \right]^{-N},
\label{16}
\end{equation}
where $2p_{\rm F}r_c/\hbar = (18 \pi Z)^{1/3} \eta$,
$N= 3.781 - 0.02322 \, Z$,
and the functions $F_0$ and  $F_1$ are given by
\begin{eqnarray}
       F_0  &  =  &   \frac{1}{\alpha}
                    \left( {\rm e}^{-u} - {\rm e}^{-\alpha} \right)
                    -  0.6449 \,
                    {{\rm e}^{- 0.3462 \alpha} \over 1 + \alpha}
\nonumber          \\
            &  -  &    {s^2 \over 2} \,
                      \left[ \ln(s) - 0.5 \right] \,
                        {\rm e}^{- \alpha}
                      +    {\alpha (1+u) \over (1 + \alpha)^3} \,
                   \left[ \ln \left(\frac{1 + u}{1+ \alpha} \right)
                   - 0.4230 \right]  \,
                         {\rm e}^{-u} ,
\label{17} \\
           F_1  & = & \frac{1}{\alpha^2} \left[ (1+u) \, {\rm e}^{-u}
                     - (1+ \alpha) \, {\rm e}^{-\alpha} \right]
                     - 0.3949 \,
                     \frac{ {\rm e}^{-0.4726 \alpha} }{(1.5 + \alpha)^{3/4}}
\nonumber         \\
                & - &
                   \frac{s^3}{3} \, \left[ \ln(s) - 5.213 \right]
                    {\rm e}^{- \alpha}
                +  \frac{\alpha \, (2+4u+u^2)}{(1.5 + \alpha)^4}
                \left[ \ln \left( \frac{1+u}{1.5 + \alpha} \right) -
                0.1 \right] \, {\rm e}^{-u}.
\label{18}
\end{eqnarray}
Here, $s = (4Z)^{-2/3}+0.00563$, $ u= s \, \alpha$.
The mean error of the fit (\ref{16}) is 1.3 \%,
and the maximum error of 3.9 \% takes place at $t$=0.12, $Z$=60,
$\eta$=0.3.

Note a simple asymptote $L \approx 0.5$ which is valid in the
limiting case when $t \gg 1$, $\alpha \gg 1$ and
$\alpha y_0^2 \ll 1$. However this asymptote cannot be actually
realized in NS matter for the conditions of study (Section 2).
In the opposite case, when $t \ll 1$, we obtain
$L= \alpha_0 t^2 u_{-2} P/a_0$, where $P$ should be calculated
at $\alpha= 0.5 \alpha_0 u_{-1}$ (see Equation (\ref{7})).

% Sec.5 ***********************************************************
\section{Discussion}
%******************************************************************
Equations (\ref{4}), (\ref{13}), (\ref{15}) and
(\ref{16}) allow us to evaluate easily the neutrino energy
loss rate $Q$ due to electron--phonon scattering for any
model of stellar matter in the parameter range of study
(Section 2).

Figures 2 -- 4 display $L$ as a function of the dimensionless
temperature $t$. The maximum values of $t$ correspond to the melting
temperature (\ref{3}). Following Itoh et al. (1984a),
the proton core radius $r_c$ in the nuclear formfactor
(\ref{9}) has been taken as:
$ r_c = 1.15 A^{1/3}$ fm at
$ \rho < 4 \times 10^{11}$ g~cm$^{-3}$; and
$ r_c = 1.83 Z^{1/3}$ fm at higher $\rho$.
If $\rho \la 10^{11}$ g~cm$^{-3}$ the
nuclear size effect is unimportant and one can set $\eta=0$.

Figure 2 corresponds to the ground--state matter
(Section 2) at $\rho = 4 \times 10^{12}$ g~cm$^{-3}$
($A=159$, $Z=50$, Negele and Vautherin, 1973).
The nuclear size effect is seen to decrease $L$ and $Q$ by (20--30) \%.
According to Equations (\ref{9}) and (\ref{14}),
the nuclear formfactor decreases the integrand
at large $y \approx 1$ (i.e., at large--angle scattering),
and the effect is more pronounced with increasing $\eta$.
However the integrand is already small at these
$y$ due to the Debye--Waller factor (which reduces the scattering
with large momentum transfers) and due to the factor
$[1+(2y^2/(1-y^2))\ln(y)]$. The latter factor describes
suppression of backward scattering of relativistic electrons
(see Haensel et al., 1996) and vanishes at $y=1$.
This explains weak dependence of $L$ on $\eta$.

The temperature dependence of $L$ in the bcc and fcc crystals
is similar. The $L(t)$ curve for the bcc crystal has formally
the maximum, just as the bcc curve, but this happens at higher
$t$, above the melting temperature.
The maximum of $L(t)$ for the fcc crystal is shifted to lower
$t$ (and seen in Figure 2) owing to a softer phonon spectrum
(see Baiko and Yakovlev, 1995).

Let us emphasize that the neutrino emission due to the
electron--phonon scattering in the bcc crystal was calculated
earlier by Itoh et al. (1984a). These results are also included
in the new review article of Itoh et al. (1996).
The authors calculated the function $F_{\rm phonon}$
related to our function $L$ as $F_{\rm phonon} = 2L/3$.
The calculations of $F_{\rm phonon}$ were carried out for the nuclei of
$^4$He, $^{12}$C, $^{16}$O, $^{20}$Ne, $^{24}$Mg,
$^{28}$Si, $^{32}$S, $^{40}$Ca, $^{56}$Fe
at 10$^4$ g~cm$^{-3} \la \rho \la 10^{12}$ g~cm$^{-3}$
and also for the ground--state matter
at 10$^4$ g~cm$^{-3} \la \rho \la 10^{14}$ g~cm$^{-3}$.
The results were fitted (Itoh et al. 1984a, 1996)
by very complicated expressions which
require large tables of fit parameters.
Our fits (Section 4) are much simpler and valid for any nuclear
composition. For comparison, Figure 2 presents
the temperature dependence of
$L(t)$ (long dashes) for the bcc crystal derived using the fits
of Itoh et al. (1984a, 1996). The results are somewhat different
from those obtained in the present article.
The difference (which is sometimes
rather large) takes place also for other values of
$\rho$, $T$, and for other elements (Figures 3 and 4).

This difference is likely to come from not
too accurate fits of Itoh et al. (1984a, 1996).
In order to confirm this statement let us
compare the equations of
Itoh et al. (1984a) and the equations
of the present article. Itoh et al. (1984a)
used another method of calculation taken from the
earlier article of Flowers (1973). For comparison,
we express the integral (\ref{14}) in Equation (\ref{13})
as a three-dimensional integral over electron momentum transfers
d$\vect{q}$. Flowers (1973) and Itoh et al. (1984a)
replaced the latter integration by summation over possible
inverse lattice cells where $\vect{q}$ could appear.
The integration over any cell of the sum was
done by replacing $\vect{q} = \vect{K}$.
In addition, just as in the present article, the normal
scattering processes were neglected, i.e., the contribution
from the central cell was excluded.
For comparison, we have also calculated $L$
by the same method of summation over inverse lattice cells.
The results (short-dashed curves in Figures 3 and 4)
are in good agreement with our original results
(for chosen nuclei and densities) but
differ from those given by the fits of Itoh et al.
(1984a, 1986) by a factor of several.

Figure 5 demonstrates the neutrino energy loss
rate $E=Q/\rho$ per unit mass
(ergs s$^{-1}$ g$^{-1}$) as a function of density
at several values of temperature for the ground--state
and accretion matters (Section 2). Note that the selfconsistent
models of accreted matter correspond to temperatures
$T \sim 10^8$~K (Miralda-Escud$\acute{e}$  et al., 1990).
We have extended the accretion model to higher $T$
for illustrative purpose, to demonstrate the effect of
variation of nuclear composition onto the neutrino emission rate.
For the displayed parameters, the matter can be
either in the liquid or in the solid
phase. The emissivity $E$ in the solid phase has been calculated
using the expressions of Section 4 (for bcc crystals).
In the liquid phase, we have used the formulae of Haensel et al. (1996).
Nuclear composition varies at certain values of density
as described in Section 2 (Figure 1). In addition, there
occur liquid-solid phase transitions at certain $\rho$
which are easily deduced from Figure 1.
For all $T$ displayed in Figure 5, the
matter of sufficiently high density
becomes solid. With decreasing $T$, the crystallization boundary
shifts towards lower densities. The accreted matter solidifies
at somewhat higher $\rho$ since it contains nuclei with lower $Z$.
In addition, if $T= 8 \times 10^8$~K
and $\rho$ increases in the accreted matter
and becomes higher than $\rho \ga 10^{11}$ g~cm$^{-3}$
there occurs a series of transitions from liquid into solid and back.
This effect is associated with strongly nonmonotonic melting temperature
due to rapid changes of $Z$ (Figure 1).

According to Equation (\ref{4}),
the dependence of the neutrino energy loss rate
$E$ on $\rho$ and $T$ is determined by the factor
$(X_A Z^2/A) \, T_8^6 \, L$. If
$X_A Z^2/A$ and $L$ were fixed, the emissivity $E$ would be displayed
by horizontal lines in Figure 5. Actually, however,
$E$ varies with $\rho$. The curves are seen to suffer jumps
associated with changes of the nuclear composition and
crystallization. The jumps due to variations of the nuclear
composition are much weaker. In a crystallization point,
the function $L$ and the emissivity $E$
decrease steeply by a factor of 1.5--3
but the general character of the density dependence of
$E$ in the liquid phase and just after the crystallization
is the same. This is because
the neutrino--pair bremsstrahlung by electrons
due to the Coulomb scattering in a liquid is similar
to that due to the high-temperature electron--phonon scattering
($T \ga T_{\rm p}$). Indeed, the function $L$ is rather smooth
either in the liquid (Haensel et al., 1996),
or in the high--temperature crystal, and it behaves as
$E \propto (X_A Z^2/A) T^6$, in the first approximation.
The crystallization jump of $E$ can be somewhat smoothed
if we additionally take into account the exponentially suppressed
static lattice contribution (Pethick and Thorsson, 1994).
However, our preliminary estimates based on the approximate expressions
given by Pethick and Thorsson (1994) and
Itoh et al. (1996) show that the smoothing is not significant.

With increasing $\rho$ inside the crystalline phase
(at a fixed temperature $T$), the ratio
$t=T/T_{\rm p}$ generally decreases, and the low--temperature
phonon scattering becomes important. In this regime,
according to the results of Section 4, the function
$L= \alpha_0 t^2 u_{-2} P/a_0$, and $P$
depends weakly on the parameters of stellar matter.
Then one has a strong ($\sim t^2$) suppression of the
neutrino generation. The generation rate becomes a sharper
function of $T$ and depends explicitly on $\rho$.
In the first approximation, $E \propto T^8
(A/Z)^{1/3} \, X_A^{1/6} \, \rho^{-5/6}$.
The suppression leads to a noticeable decrease of
$E$ with $\rho$ (Figure 5). The dependence of $E$ on nuclear composition
becomes much weaker, than in the liquid or high--temperature
crystal. This explains increasing similarity of the
emissivities $E(\rho)$ for the ground--state and accreted matters
in Figure 5 with decreasing $T$ and/or increasing $\rho$. At
$T \la 5 \times 10^8$~K the difference of the emissivities
becomes negligibly small.

As follows from Figure 5, the neutrino emissivity
$E$ is much more sensitive to the state of matter
(liquid or crystal) than to nuclear composition
(accreted or ground--state matter). For instance, it is easy
to calculate the emissivity $E$ for matter composed of iron nuclei
$^{56}$Fe and show that it differs slightly from the
curves presented in Figure 5.

% Sec. 6 **********************************************************
\section{Conclusions}
%******************************************************************
We have calculated (Section 3) the neutrino energy loss rate
due to electron--phonon scattering in the Coulomb bcc and
fcc crystals of spherical atomic nuclei at densities
$10^6$ g~cm$^{-3}  \ll  \rho  \la 10^{14}$ g~cm$^{-3}$
and temperatures $T_{\rm U} \la T \leq T_{\rm m}$
in matter of arbitrary nuclear composition (Section 2).
The results are fitted by simple expressions (Section 4).
Combined with the results of Haensel et al. (1996),
which describe the neutrino emission due to
the bremsstrahlung in the liquid phase
($T_{\rm m} < T \la T_{\rm F}$), the equations obtained
allow one to evaluate easily the neutrino energy loss rate
due to the bremsstrahlung process in the wide temperature range
$T_{\rm U} \la T \la T_{\rm F}$, most important for applications.
The main properties of the neutrino energy losses
are analysed in Section 5 for two models of matter
in a NS core (described in Section 2):
ground--state and accreted matter. Note that at very low
temperature ($T \ll T_{\rm m}$) the neutrino emission
due to the electron--phonon scattering may appear so weak
that another neutrino generation mechanism
becomes dominant --- the Coulomb scattering of electrons by
charged impurities. The corresponding neutrino energy loss rate
can be easily calculated, for instance, from the results
of Haensel et al. (1996). However the neutrino emission
at these low temperatures is expected to be
too weak to be of practical interest.

The results of the present article can be useful for numerical
modelling of various processes associated with thermal evolution
of NSs. First of all, we mean cooling of young NSs
(of age $\la (1$ -- $10^3)$ yrs), where thermal relaxation
is not yet achieved (Lattimer et al., 1994).
The relaxation is accompanied by a
cooling wave which goes from the stellar interior to the surface.
Corresponding variations of the surface temperature
are, in principle, observable. The dynamics of thermal
relaxation is quite sensitive to the properties of matter
in the NS crust, especially to nuclear composition and
neutrino energy losses.\\[2ex]

The authors are grateful to C. Pethick and V. Thorsson
for very useful discussions,
to D.A. Baiko for assistance in Monte Carlo calculations, and
to N. Itoh for presenting the review article of
Itoh et al. (1996) prior to publication.
This work was partially supported
by the Russian Basic Research Foundation
(grant 93-02-2916), the Soros Foundation (grant R6A-003),
and INTAS (grant 94-3834).

\newpage
\begin{center}
                   {\bf References}
\end{center}

\noindent
Baiko, D.A. and Yakovlev, D.G., Pisma Astron. Zh. (Astron. Lett.),
          1995, v. 21, p. 784.

\noindent
Brush, S.G., Sahlin, H.L., Teller, E. J., Chem. Phys., 1966, v. 45, p. 2102.

\noindent
Carr, W.J., Phys. Rev., 1961, v. 122, p. 1437.

\noindent
Flowers, E., Astrophys. J., 1973, v. 180, p. 911.

\noindent
Haensel, P. and Zdunik, J.L., Astron. Astrophys., 1990a, v. 227, p. 431.

\noindent
Haensel, P. and Zdunik, J.L., Astron. Astrophys., 1990b, v. 229, p. 117.

\noindent
Haensel, P. and Pichon, D., Astron. Astrophys., 1994, v. 283, p. 313.

\noindent
Haensel, P., Kaminker, A.D., Yakovlev, D.G., Astron. Astrophys.,
    1996 (in press).

\noindent
Ito, N. and Kohyama, Y., Astrophys. J., 1983, v. 275, p. 858.

\noindent
Itoh, N., Kohyama, Y., Matsumoto, N., Seki, M.,
   Astrophys. J., 1984a, v. 285, p. 304.

\noindent
Itoh, N., Kohyama, Y., Matsumoto, N., Seki, M.,
   Astrophys. J., 1984b, v. 285, p. 758; erratum: v. 404, p. 418.

\noindent
Itoh, N., Adachi, T., Nakagawa, M., Kohyama, Y., Munakata, H.,
   Astrophys. J., 1989, v. 339, p. 354; erratum: 1990, v. 360, p. 741.

\noindent
Itoh, N., Hayashi, H., Kohyama, Y., Astrophys. J., 1993, v. 418, p. 405.

\noindent
Itoh, N, Hayashi, H., Nishikawa, A., Kohyama, Y., Astrophys. J. Suppl.,
    1996 (in press).

\noindent
Jancovici, B., Nuovo Cimento, 1962, v. 25, p. 428.

\noindent
Lattimer, J.M., Van Riper, K., Prakash, M., Prakash, M., Astrophys. J.,
    1994, v. 425, p. 802.

\noindent
Lorenz, C.P., Ravenhall, D.G., Pethick, C.J., Phys. Rev. Lett., 1993,
    v. 70, p. 379.

\noindent
Miralda-Escud$\acute{e}$, J., Haensel, P., Paczy$\acute{n}$ski, B.,
    Astrophys. J., 1990, v. 362, p. 572.

\noindent
Mochkovitch, R. and Hansen, J.-P., Phys. Lett., 1979, v. A73, p. 35.

\noindent
Nagara, H., Nagata, Y., Nakamura, T., Phys. Rev., 1987, v. A36, p. 1859.

\noindent
Negele, J.W. and Vautherin, D., Nucl. Phys., 1973, v. A207, p. 298.

\noindent
Pethick, C.J. and Ravenhall, D.G., Ann. Rev. Nucl. Particle Sci., 1995,
    v. 45, p. 429.

\noindent
Pethick, C.J. and Thorsson, V., Phys. Rev. Lett., 1994, v. 72, p. 1964.

\noindent
Raikh, M.E. and Yakovlev, D.G., Astrophys. Space Sci., 1982, v. 87, p. 193.

\noindent
Yakovlev, D.G. and Urpin, V.A., Sov. Astron. Lett., 1980, v. 24, p. 303.

\noindent
Ziman, J.M., Electrons and Phonons, Oxford University Press: Oxford, 1960.

\newpage

\begin{figure}[t]
\epsfxsize=0.8\hsize
\centerline{{\epsfbox{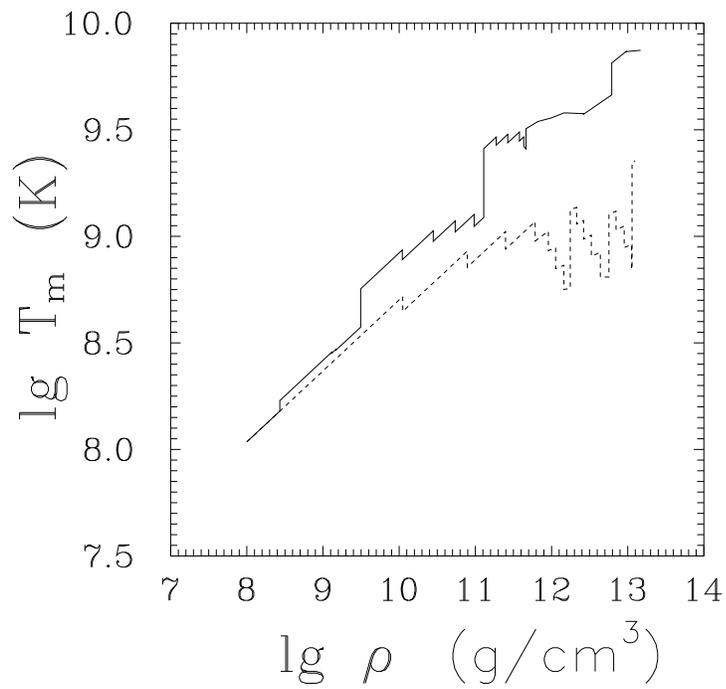}}}
%\vskip 6.5truecm
\caption{\label{figure1}
Melting temperatures of the ground--state (solid line)
and accreted (dash line) matter.
}
\end{figure}

\newpage
\begin{figure}[t]
\epsfxsize=0.8\hsize
\centerline{{\epsfbox{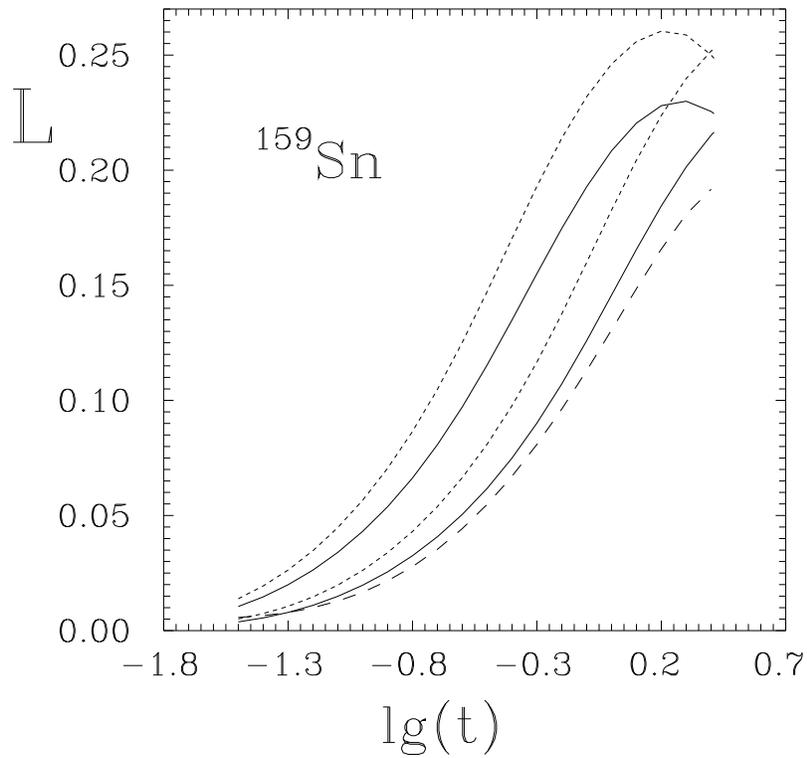}}}
%\vskip 6.5truecm
\caption{\label{figure2}
Quantity $L$ versus $t=T/T_{\rm p}$
for the bcc (lower curves)
and fcc (upper curves) crystals composed of
$^{159}$Sn nuclei at density
$\rho= 4 \times 10^{12}$ g~cm$^{-3}$ with account for the
finite nuclear size ($\eta=0.15$, solid lines)
and for point-like nuclei ($\eta=0$, short dashes).
Long dashes --- fits of Itoh et al. (1984a).
}
\end{figure}

\newpage

\begin{figure}[t]
\epsfxsize=0.8\hsize
\centerline{{\epsfbox{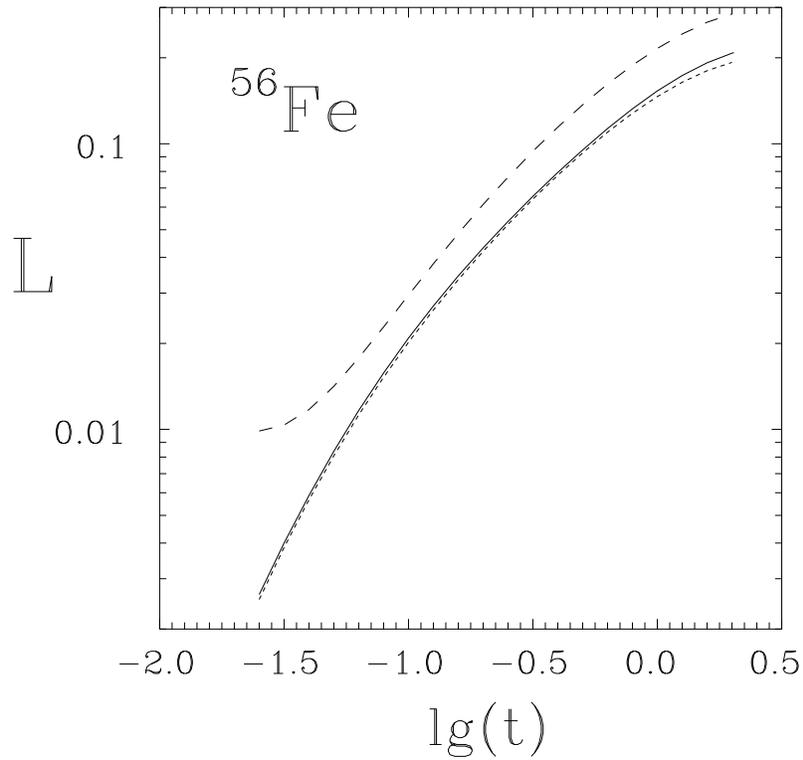}}}
%\vskip 6.5truecm
\caption{\label{figure3}
Quantity $L$ versus $t$ for the bcc crystal composed of
$^{56}$Fe nuclei at $\rho = 10^9$ g~cm$^{-3}$. Solid line ---
present calculation, long dashes --- fit of
Itoh et al. (1984a, 1996), short dashes ---
present calculation using the method of Itoh et al. (1984a).
}
\end{figure}

\newpage

\begin{figure}[t]
\epsfxsize=0.8\hsize
\centerline{{\epsfbox{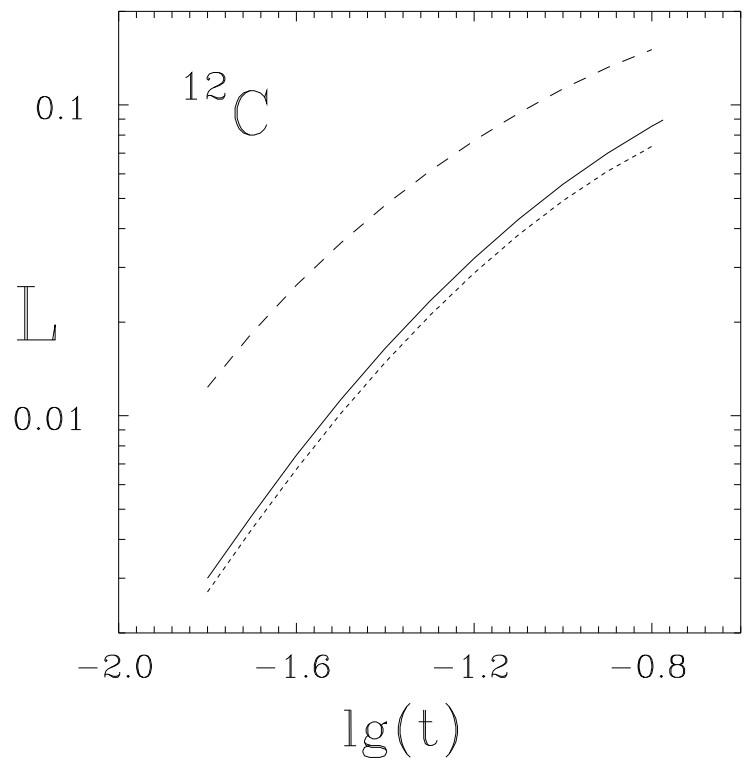}}}
%\vskip 6.5truecm
\caption{\label{figure4}
Same as in Figure 3, but for the crystal composed of $^{12}$C.
}
\end{figure}

\newpage

\begin{figure}[t]
\epsfxsize=0.8\hsize
\centerline{{\epsfbox{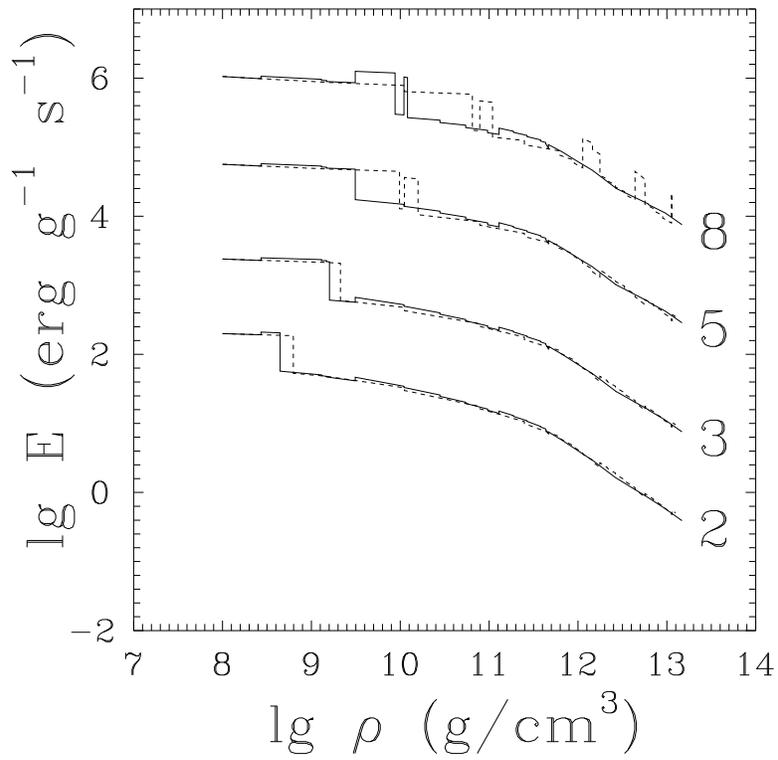}}}
%\vskip 6.5truecm
\caption{\label{figure5}
Neutrino energy loss rate due to the bremsstrahlung produced by
electron-nucleus ($T>T_{\rm m}$) or electron--phonon scattering
($T<T_{\rm m}$) versus density for the ground--state
(solid lines) and accreted (dash lines) matters at several values
of $T$ (values of $T_8$ are shown near the curves).
}
\end{figure}

\end{document}